\documentclass[twocolumn,showpacs,amsmath,amssymb,letter]{revtex4}
\usepackage{epsfig}
\usepackage{bm}
\usepackage{graphicx}
\usepackage{dcolumn}% Align table columns on decimal point

\begin{document}

\title{Direct Assessment of Vorticity Alignment with Local and Nonlocal
Strain Rates in Turbulent Flows}
\author{Peter E. Hamlington$^1$,
         J\"org Schumacher$^2$, and
         Werner J.A. Dahm$^1$}
  \affiliation{$^1$ Laboratory for Turbulence \& Combustion (LTC),
  Department of Aerospace Engineering, The University of Michigan, Ann
Arbor, MI 48109-2140, USA\\
  $^2$ Institute for Thermodynamics and Fluid Mechanics, Technische Universit\"at
Ilmenau, D-98684 Ilmenau, Germany}

\date{\today}

\begin{abstract}

A direct Biot-Savart integration is used to decompose the strain
rate into its local and nonlocal constituents, allowing the
vorticity alignment with the local and nonlocal strain rate
eigenvectors to be investigated.  These strain rate tensor
constituents are evaluated in a turbulent flow using data from
highly-resolved direct numerical simulations.  While the vorticity
aligns preferentially with the intermediate eigenvector of the
\textit{combined} strain rate, as has been observed previously,
the present results for the first time clearly show that the
vorticity aligns with the most extensional eigenvector of the
\textit{nonlocal} strain rate. This in turn reveals a significant
linear contribution to the vortex stretching dynamics in turbulent
flows.

\end{abstract} \noindent

\maketitle\pagebreak

The alignment of vorticity with the strain rate eigenvectors in
turbulent flows has been a subject of considerable interest over
the past two decades. Since the initial finding \cite{ashurst1987}
that the vorticity shows a preferred alignment with the
intermediate eigenvector of the strain rate tensor, there have
been numerous studies seeking to understand the reasons for this
result, and various theoretical approaches have been proposed to
explain the failure of the vorticity to align with the most
extensional strain rate eigenvector.

In this Letter, we help resolve this issue by showing that
vorticity in turbulence does tend toward alignment with the most
extensional eigenvector of the \textit{nonlocal} (background)
strain, namely the strain field induced in the immediate region
around any vortical structure by the surrounding vorticity outside
this region. The anomalous alignment occurs with the eigenvectors
of the \textit{combined} strain rate, namely the sum of this
nonlocal background strain and the \textit{local} strain induced
in the region by the vorticity within it.

Alignment of the vorticity vector $\bm{\omega}\equiv \nabla
\times\textbf{u}$ with the strain rate tensor $S_{ij}$ in
three-dimensional incompressible turbulent flows is ultimately
responsible for the transfer of kinetic energy between scales,
and for the nonlinearity in the dynamics of the
underlying vorticity field.  The inverse curl operator is
the Biot-Savart integral that gives the velocity field
$\textbf{u}$ from the vorticity field $\bm{\omega}$ as
\begin{equation}\label{v1}
   \textbf{u}(\textbf{x}) = \frac{1}{4\pi}
   \int_{\textbf{x}^\prime}\bm{\omega}(\textbf{x}^\prime)\times
   \frac{\textbf{x}-\textbf{x}^\prime}{|\textbf{x}-\textbf{x}^\prime|^3}
   d^3 \textbf{x}^\prime\,.
\end{equation}
The resulting gradients of $\textbf{u}$ define the strain rate
tensor $S_{ij} = \frac{1}{2}\left(\partial u_i/\partial x_j
+\partial u_j/\partial x_i\right)$ which, in turn, is coupled to
the dynamics of the vorticity as
\begin{equation}\label{v3}
   \frac{D\omega_i}{Dt} = S_{ij}\omega_j + \nu \nabla^2 \omega_i\,.
\end{equation}
On the right side of (\ref{v3}), the magnitude of the stretching
term $|S_{ij}\omega_j| \equiv \omega [s^2_i\left(\textbf{e}_i\cdot
\textbf{e}_\omega\right)^2]^{1/2}$ depends on the strain rate
eigenvalues $s_i$ and the vorticity magnitude $\omega\equiv
\left(\omega_i \omega_i\right)^{1/2}$, and on the alignment
cosines $\left(\textbf{e}_i\cdot \textbf{e}_\omega\right)$ between
the vorticity unit vector $\textbf{e}_\omega$ and the strain rate
eigenvectors $\textbf{e}_i$.

The strain rate eigenvalues $s_i$ can be ordered as $s_1\geq s_2
\geq s_3$, so that incompressibility
$\left(s_1+s_2+s_3\equiv0\right)$ requires $s_1\geq 0$ and
$s_3\leq 0$. The positivity of $s_1$ and the negativity of $s_3$
correspond, respectively, to extensional and compressional
straining along the $\textbf{e}_1$ and $\textbf{e}_3$ directions.
While the intermediate eigenvalue $s_2$ is on average weakly
positive in turbulent flows, the instantaneous $s_2$ can take on
large positive or negative values
\cite{ashurst1987,tsinober2001,su1996} bounded only by the $s_1$
and $s_3$ values.  The alignment between the vorticity and the
strain rate eigenvectors similarly determines the production rate
$\omega_i S_{ij}\omega_j \equiv \omega^2 s_i
\left(\textbf{e}_i\cdot \textbf{e}_\omega\right)^2$ for the
enstrophy $\frac{1}{2}\left(\bm{\omega}\cdot\bm{\omega}\right)$.
The three alignment cosines $\left(\textbf{e}_i\cdot
\textbf{e}_\omega\right)$ thus play an essential role in the
structure and dynamics of turbulent flows.

Despite its importance, the mechanism by which the vorticity
aligns with the strain rate eigenvectors $\textbf{e}_i$ is still
not well understood.  In particular, the maximality and positivity
of $s_1$ might suggest that the vorticity in (\ref{v3}) would show
preferred alignment with the most extensional eigenvector
$\textbf{e}_1$. However, since $S_{ij}$ on the right side of
(\ref{v3}) is coupled back to $\omega_i$, the resulting
nonlinearity complicates any such simple alignment. Indeed,
numerous studies have shown that the vorticity in turbulent flows
instead shows a preference for alignment with the intermediate
strain rate eigenvector $\textbf{e}_2$.

\begin{figure}[t]
\centering {\includegraphics[width=3.1in]{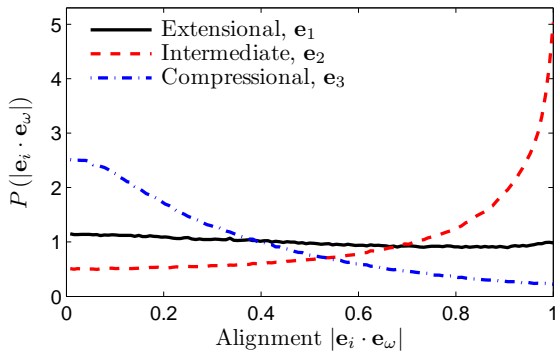}} \caption{(Color
online) Distributions of alignment cosines $|\textbf{e}_i\cdot
\textbf{e}_\omega|$.} \label{f1}
\end{figure}

This can be seen, for example, in Figure \ref{f1}, where
distributions of the three alignment cosines $|\textbf{e}_i\cdot
\textbf{e}_\omega|$ are shown from a recent highly-resolved,
three-dimensional, $2048^3$ direct numerical simulation (DNS)
\cite{schumacher2007} of statistically stationary, forced,
homogeneous, isotropic turbulence at Taylor-scale Reynolds number
$Re_\lambda =107$. The simulation was done using a pseudospectral
method with a spectral resolution that exceeds the standard value
by a factor of eight. As a result the highest wavenumber
corresponds to $k_{max}\eta_K=10$, and the Kolmogorov lengthscale
$\eta_K$ is resolved with three grid spacings. The resulting
alignment distributions in Fig.\ \ref{f1} agree with those from
lower-resolution DNS studies as well as from laboratory
measurements \cite{su1996,mullin2006,tsinober1992}. In particular,
the vorticity tends to point away from the most compressive
eigenvector $\textbf{e}_3$, namely $|\textbf{e}_3\cdot
\textbf{e}_\omega|\rightarrow 0$, and there is essentially no
tendency for any preferred alignment relative to the most
extensional strain rate eigenvector $\textbf{e}_1$, since
$P\left(|\textbf{e}_1\cdot \textbf{e}_\omega|\right)\approx 1$.
However, the vorticity shows a strong tendency toward alignment
with the intermediate strain rate eigenvector $\textbf{e}_2$,
namely $|\textbf{e}_2\cdot \textbf{e}_\omega|\rightarrow 1$.

Previous attempts to understand these alignments have often
focused on the vorticity-strain coupling in the evolution equation
for the strain rate tensor, namely
\begin{equation}\label{v4}
   \frac{DS_{ij}}{Dt} = -S_{ik}S_{kj} - \frac{1}{4}\left(\omega_i
   \omega_j - \omega_k \omega_k \delta_{ij}\right) - \frac{1}{\rho}
   \Pi_{ij} + \nu \nabla^2 S_{ij}\,,
\end{equation}
where $\Pi_{ij} \equiv \partial^2 p/\partial x_i \partial x_j$ is
the pressure Hessian. The nonlinear coupling between the strain
rate and the vorticity is apparent in (\ref{v3}) and (\ref{v4}),
where the nonlocality of the vorticity and strain rate
co-evolution involves the pressure Hessian
\cite{ohkitani1995,nomura1998}. While progress has been made in
understanding the vorticity alignments in Fig.\ \ref{f1} via the
restricted Euler equations (e.g.\ \cite{nomura1998}), where
nonlocal effects are neglected entirely, a complete picture that
clearly distinguishes between local and nonlocal contributions to
the vorticity dynamics has remained largely elusive.

Here, we forgo the use of (\ref{v4}) in representing the strain
rate appearing in (\ref{v3}), instead using an integral
representation for the strain rate derived from (\ref{v1}), and
then use this to gain insights into the effects of nonlocality and
nonlinearity on the vorticity alignment.  As shown in
\cite{hamlington2008}, from the Biot-Savart integral in (\ref{v1})
the strain rate can be exactly expressed in terms of the vorticity
as
\begin{equation}\label{v5}
   S_{ij}(\textbf{x}) = \frac{3}{8\pi} \int_{\textbf{x}^\prime}
   \left( \epsilon_{ikl}r_j + r_i \epsilon_{jkl}\right)
   \frac{r_k}{r^5} \omega_l (\textbf{x}^\prime) d^3
   \textbf{x}^\prime\,,
\end{equation}
where $\textbf{r}\equiv \textbf{x}-\textbf{x}^\prime$, $r\equiv
|\textbf{r}|$, and the integral is defined in a principal value
sense. Substituting (\ref{v5}) in (\ref{v3}) then gives a direct
nonlocal integro-differential equation for the vorticity evolution
as
\begin{equation}\label{v6}
   \frac{D\omega_i}{Dt} = \frac{3}{8\pi}\int_{\textbf{x}^\prime}
   \left( \epsilon_{ikl}r_j + r_i \epsilon_{jkl}\right)
   \frac{r_k}{r^5} \left[\omega_j (\textbf{x})\omega_l
(\textbf{x}^\prime)\right] d^3
   \textbf{x}^\prime + \nu \nabla^2\omega_i\,,
\end{equation}
which depends only on the vorticity field itself. In (\ref{v5})
and (\ref{v6}), the local and nonlocal contributions to the strain
rate and vorticity dynamics can be understood by separating the
integration domain into a \textit{local} region of radius $r \leq
R$ centered on $\textbf{x}$, and a \textit{nonlocal} region that
accounts for the rest of the domain \cite{hamlington2008,she1991}.
The strain rate in (\ref{v5}) then is the sum
\begin{equation}\label{v7}
   S_{ij}(\textbf{x}) = S^R_{ij}(\textbf{x})+
   S^B_{ij}(\textbf{x})
\end{equation}
of the local strain rate $S^R_{ij}(\textbf{x})$ induced at
$\textbf{x}$ by the vorticity within $R$, and the nonlocal
(background) strain rate $S^B_{ij}(\textbf{x})$ induced at
$\textbf{x}$ by all the vorticity outside $R$.

The background strain field $S^B_{ij}(\textbf{x})$ in the vicinity
of any local vortical structure in the turbulence is that induced
by all the \textit{other} vortical structures. Thus, the proper
physical value for $R$ used to obtain $S^B_{ij}(\textbf{x})$
should exclude from (\ref{v5}) essentially all the vorticity
associated with any local vortical structure. Prior studies (e.g.\
\cite{jimenez1993}) have shown that the characteristic radius of
intense vortical structures in turbulence is in the range
$r/\eta_K\approx 4-10$, where $\eta_K$ is the Kolmogorov length
scale. This is consistent with the two-point vorticity correlation
from the present DNS of homogeneous isotropic turbulence, which is
found to decrease to 20\% of its maximum value at
$r\approx12\eta_K$. This gives a physically appropriate cutoff
radius, since beyond this the vorticity becomes essentially
uncorrelated with itself.  Thus, $R= 12\eta_K$ as used herein
excludes essentially all the local vorticity for most structures,
and thereby allows the self-induced strain field in the vicinity
of typical vortical structures to be separated from the background
strain field in which the structures reside.

Although \cite{hamlington2008} developed an operator for the
background strain rate $S^B_{ij}(\textbf{x})$, here we avoid the
associated infinite Taylor series with respect to $R$ and instead
obtain the local strain rate $S^R_{ij}(\textbf{x})$ by directly
integrating (\ref{v5}) over the domain $R$ centered on
$\textbf{x}$. At any point $\textbf{x}$ in the $2048^3$ cubic
simulation domain, a smaller cubic subdomain with side length $2R$
is taken to define the local region around $\textbf{x}$. The local
strain rate $S_{ij}^R(\textbf{x})$ is obtained by numerically
integrating (\ref{v5}) over this subdomain. We then determine the
nonlocal strain rate from (\ref{v7}) as $S^B_{ij}(\textbf{x}) =
S_{ij}(\textbf{x})-S^R_{ij}(\textbf{x})$, and examine the
alignment of the vorticity $\bm{\omega}(\textbf{x})$ with each of
these strain rates to understand how the alignment in Fig.\
\ref{f1} arises.

\begin{figure}[t]
\centering {\includegraphics[width=3.1in]{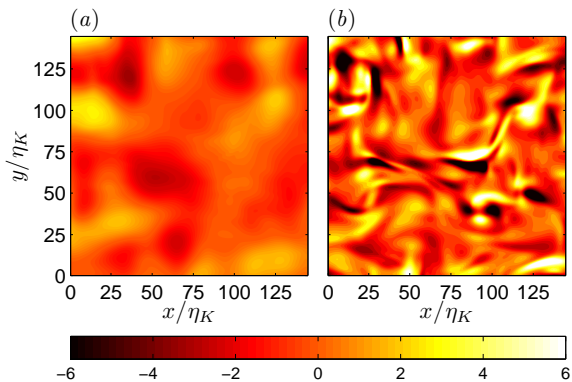}} \caption{(Color
online) Typical example of (\textit{a}) background strain rate
field $S^B_{12}(\textbf{x})$ and (\textit{b}) local strain rate
field $S^R_{12}(\textbf{x})$.} \label{f3}
\end{figure}

Figure \ref{f3} shows an example of the resulting decomposition of
the shear strain rate field $S_{12}(\textbf{x})$ into its
background and local fields, $S^B_{12}(\textbf{x})$ and
$S^R_{12}(\textbf{x})$. Similar local-nonlocal decompositions are
obtained for the other strain rate components, and the eigenvalues
and eigenvectors of the resulting background and local strain rate
tensor fields are then computed. At every point $\textbf{x}$, the
alignment cosines $|\textbf{e}_i\cdot \textbf{e}_\omega|$ of the
vorticity with the background and local strain rate eigenvectors,
denoted $\textbf{e}^B_i$ and $\textbf{e}^R_i$ respectively, are
then evaluated.

\begin{figure}[b]
\centering {\includegraphics[width=3.1in]{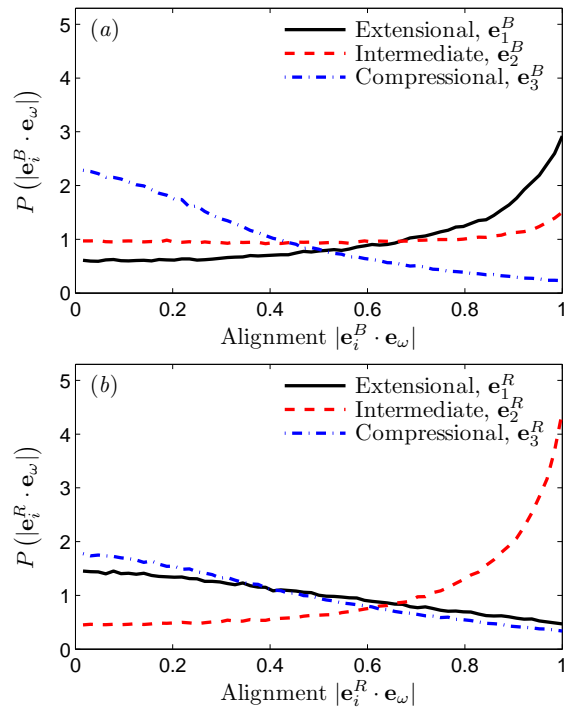}} \caption{(Color
online) Distributions of vorticity alignment cosines for
(\textit{a}) background strain rate eigenvectors $\textbf{e}^B_i$,
and (\textit{b}) local strain rate eigenvectors $\textbf{e}^R_i$.}
\label{f4}
\end{figure}

The resulting vorticity alignment distributions
$P\left(|\textbf{e}^B_i \cdot \textbf{e}_\omega|\right)$ and
$P\left(|\textbf{e}^R_i \cdot \textbf{e}_\omega|\right)$ are
shown, respectively, in Figs.\ \ref{f4}\textit{a} and
\ref{f4}\textit{b}. From the \textit{background} strain alignments
in Fig.\ \ref{f4}\textit{a}, it is apparent that the vorticity is
preferentially aligned with the most \textit{extensional}
background strain rate eigenvector $\textbf{e}^B_1$, namely
$|\textbf{e}^B_1\cdot \textbf{e}_\omega|\rightarrow 1$. There is
essentially no preferred alignment of the vorticity relative to
the intermediate background eigenvector $\textbf{e}^B_2$, since
$P\left(|\textbf{e}^B_2\cdot \textbf{e}_\omega|\right) \approx 1$,
while the vorticity tends to point preferentially away from the
most compressive background eigenvector $\textbf{e}^B_3$, namely
$|\textbf{e}^B_3\cdot \textbf{e}_\omega|\rightarrow 0$.

The alignments in Fig.\ \ref{f4}\textit{a} with the background
strain rate are precisely as would be expected when the strain
rate evolution is decoupled from that of the vorticity, as is
essentially the case for the background strain. From (\ref{v3})
with (\ref{v7}), the inviscid dynamics of the vorticity satisfies
\begin{equation}\label{v11}
   \frac{D\omega_i}{Dt} = S^B_{ij}\omega_j + S^R_{ij}\omega_j\,.
\end{equation}
By definition, the background strain rate $S^B_{ij}$ in
(\ref{v11}) is independent of the vorticity at $\textbf{x}$, and
thus its effect on the dynamics of the vorticity
$\omega_i(\textbf{x})$ is essentially linear. Since $s^B_1\geq 0$
and $s^B_3\leq 0$, and since $s_2^B\leq s^B_1$, the effect is to
cause $\bm{\omega}$ to rotate toward alignment with the most
extensional eigenvector $\textbf{e}^B_1$ of $S^B_{ij}$. The fact
that such alignment of the vorticity is seen in Fig.\
\ref{f4}\textit{a} suggests that the quasi-linear dynamics from
the first term on the right in (\ref{v11}) plays at least a
significant role in the overall evolution of the vorticity field
in turbulent flows. In the terminology of She \textit{et al.}
\cite{she1991}, this would be referred to as \textit{kinematic
nonlocality}, as distinguished from the \textit{dynamic locality}
that was the focus of their study.

From the vorticity alignments in Fig.\ \ref{f4}\textit{b} with the
\textit{local} strain rate field induced in $R$ by the local
vorticity, $\bm{\omega}$ shows substantial and essentially equal
preference for pointing largely perpendicular to the most
extensional and compressional eigenvectors $\textbf{e}^R_1$ and
$\textbf{e}^R_3$ of the local strain rate $S^R_{ij}$, namely
$|\textbf{e}^R_1\cdot \textbf{e}_\omega|\rightarrow 0$ and
$|\textbf{e}^R_3\cdot \textbf{e}_\omega|\rightarrow 0$. This is
consistent with the fact that much of the vorticity in turbulent
flows concentrates into relatively compact line-like and
sheet-like structures formed by locally axisymmetric and planar
background strain rate fields \cite{jimenez1992,buch1996}. In the
former case, the axisymmetric Burgers vortex is often used as an
idealized representation of such structures, while in the latter
case the planar Burgers vortex sheet provides a similar idealized
representation.  In both cases, the two-dimensional \textit{local}
strain field induced by the vortical structure has large
extensional and compressional eigenvalues with eigenvectors that
are necessarily perpendicular to the vorticity, due to the
geometry of the structures.  The remaining eigenvalue is zero for
perfectly two-dimensional structures, and will be nonzero only due
to small departures from strict two-dimensionality of the
structures.  Its small magnitude is thus nearly always between the
other two eigenvalues, and will therefore be the intermediate
eigenvalue.  Its eigenvector must necessarily be perpendicular to
the other two, and so will necessarily be closely aligned with the
vorticity itself.

This is precisely the alignment seen with $\textbf{e}^R_2$ in
Fig.\ \ref{f4}\textit{b}, where the vorticity points strongly
along the direction of the intermediate eigenvector, namely
$|\textbf{e}^R_2\cdot\textbf{e}_\omega|\rightarrow 1$. Note that
this `preferred' alignment of the vorticity with the intermediate
local strain eigenvector in Fig.\ \ref{f4}\textit{b} is not a
result of the nonlinear dynamics from the second term on the right
side in (\ref{v11}), but rather is a simple geometric consequence
of the largely sheet-like and line-like structures into which the
vorticity is formed \cite{jimenez1992}.

\begin{figure}[t]
\centering {\includegraphics[width=3.1in]{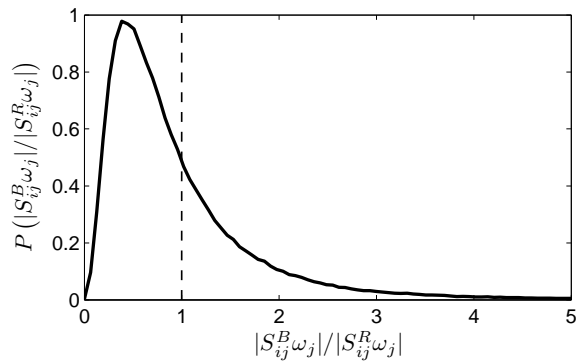}}
\caption{Distribution of background (linear) to local (nonlinear)
vortex stretching ratio.} \label{f5}
\end{figure}

The alignments in  Fig.\ \ref{f4} allow the relative contributions
from the two terms on the right side of the inviscid vorticity
dynamics in (\ref{v11}) to be understood. In particular, Fig.\
\ref{f5} shows the distribution of the ratio of background and
local vortex stretching rates, namely
$|S^B_{ij}\omega_j|/|S^R_{ij}\omega_j|$.  It is apparent that in
much of the flow this stretching ratio exceeds one, meaning that
the linear stretching dynamics produced by the background
(nonlocal) strain field $S^B_{ij}$ in (\ref{v11}) exceeds the
nonlinear stretching dynamics from the local strain field
$S^R_{ij}$. Thus, despite the overall nonlinear dynamics governing
the vorticity evolution in (\ref{v3}), a substantial part of the
underlying dynamics is linear and nonlocal.

This is consistent with the alignment in Fig.\ \ref{f4}\textit{b}
of the vorticity with the intermediate eigenvector of the local
strain rate $S^R_{ij}$, for which the associated eigenvalue
$s^R_2$ has the smallest magnitude among the three local strain
eigenvalues, and thus the associated stretching is not necessarily
large. By contrast, Fig.\ \ref{f4}\textit{a} shows that the
vorticity aligns with the most extensional eigenvector of the
background strain rate $S^B_{ij}$, and thus is stretched by the
largest of its three eigenvalues. As a result, even when
$|S_{ij}^B|$ is smaller than $|S_{ij}^R|$, the background
stretching may be larger than the local stretching. This is a
consequence of the fact that in (\ref{v11}) vortex stretching by
the local strain rate is generally not favored from the standpoint
of geometrical alignment. It is remarkable that the linear
stretching dynamics from this background (nonlocal) strain field
is comparable to the nonlinear stretching dynamics from the local
strain field.

\begin{figure}[b]
\centering {\includegraphics[width=3.1in]{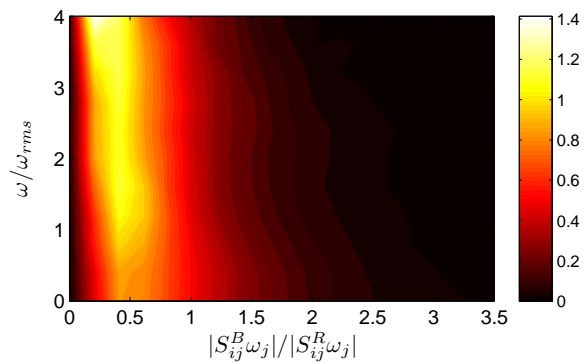}} \caption{(Color
online) Distribution of background to local vortex stretching
ratio conditioned on vorticity magnitude $\omega/\omega_{rms}$.}
\label{f6}
\end{figure}

Further insights into the background and local dynamics may be
gained by conditioning the vortex stretching ratio on the
vorticity magnitude $\omega$, as shown in Fig.\ \ref{f6}. While
there is a tendency towards smaller vortex stretching ratios for
large vorticity magnitudes in Fig.\ \ref{f6}, the observed
dependence is relatively weak. This is due to the competition
between increased local strain rate magnitude (which favors local
stretching) and the correspondence with nearly two-dimensional
intense vortical structures (which favors background stretching)
for large values of $\omega$. Unraveling the individual
contributions of these two effects, as well as consideration of
other secondary parameters such as the background strain
persistence or the Reynolds number, is an important direction for
future research. \linebreak

PH and WD acknowledge support from the Air Force Research
Laboratory (AFRL) through the Michigan-AFRL Collaborative Center
for Aeronautical Sciences (MACCAS). JS acknowledges support by the
German Academic Exchange Service (DAAD) and the Heisenberg Program
of the Deutsche Forschungsgemeinschaft (DFG). The simulations have
been carried out at the J\"ulich Supercomputing Centre (Germany).

\end{document}